\documentclass[a4paper, 10pt, twocolumn]{article}
\usepackage[utf8]{inputenc}         % Schriftkodierung dieser Datei
\usepackage[english]{babel}          % für deutsche Dokumente

\usepackage{graphicx}               % optional für Grafiken
\usepackage{tabularx}               % optional für Tabellen
\usepackage{multirow}               % optional für Tabellen
\usepackage{url}                % optional für Internet Links

\usepackage[small,bf]{caption2}     % bitte für Bildunterschriften verwenden
\usepackage{parskip}
\usepackage{titlesec}
\usepackage{amsmath}                % optional für Formeln
\usepackage{amssymb}

\titleformat{\section}{\normalfont\large\bfseries}{\thesection}{}{}
\titleformat{\subsection}{\normalfont\large\bfseries}{\thesection}{}{}
\titleformat{\paragraph}{\normalfont\bfseries}{\theparagraph}{}{}
\titlespacing{\section}{0pt}{6pt}{-1pt}
\titlespacing{\subsection}{0pt}{3pt}{-1pt}
\titlespacing{\paragraph}{0pt}{3pt}{-1pt}

\newcolumntype{Y}{>{\centering\arraybackslash}X}    %für Tabellen mit tabularx

\begin{document}

\date{}                                         % kein Datum auf 1. Seite

\title{\vspace{-8mm}\textbf{\large
Improving Voice Conversion for Dissimilar Speakers Using Perceptual Losses }}

% Hier die Namen und Daten der beteiligten Autoren eintragen
\author{
Suhita Ghosh$^1$, Yamini Sinha$^2$,  Ingo Siegert$^2$ and Sebastian
Stober$^1$\\
$^1$ \emph{\small Artificial Intelligence Lab, Otto-von-Guericke-University Magdeburg, Germany
}\\
$^2$ \emph{\small Mobile Dialog Systems, Otto-von-Guericke-University Magdeburg, Germany} } \maketitle
\thispagestyle{empty}           % weder Kopf- noch Fußzeile auf Folgeseiten
% Beginn des eigentlichen Manuskripts
\section*{Introduction}
\label{sec:Introduction} 
The rising trend of using voice as a means of interacting with smart devices has sparked worries over the protection of users' privacy and data security~\cite{wienrich2021trustworthiness}.
These concerns have become more pressing, especially after the European Union's adoption of the General Data Protection Regulation (GDPR).
The information contained in an utterance encompasses critical personal details about the speaker, such as their age, gender, socio-cultural origins and more.
If there is a security breach and the data is compromised, attackers may utilise the speech data to circumvent the speaker verification systems or imitate authorised users~\cite{RUB66}.
Therefore, it is pertinent to anonymise the speech data before being shared across devices, such that the source speaker of the utterance cannot be traced.
Voice conversion (VC) can be used to achieve speech anonymisation, which involves altering the speaker's characteristics while preserving the linguistic content.

Many voice conversion approaches have been proposed over the years, where the deep learning-based methods outperform the conventional ones~\cite{sisman2020overview}.
Further, the generative adversarial network (GAN) based approaches produce natural-sounding conversions~\cite{sisman2020overview}.
However, the quality is dependant on the selection of the target speaker.
% The GAN based approaches typically use non-parallel data, which do not require the same utterance from the source and target speakers while training.
%This is possible due to the use of consistency loss, which is calculated between the source utterance and the converted utterance conditioned on the source speaker.
% Although the overall quality of conversions produced by GANs is much better than other approaches, 
This is because GAN-based VC methods typically use non-parallel data, which prevents the computation of loss between the source utterance and the conversion conditioned on a speaker other than the source.
The quality of conversion degrades when the acoustic properties between the source and target speakers are diverse.
However, to achieve a successful anonymisation, the source and target speakers should not have very similar acoustic properties, such as pitch.

In this work, we propose perceptual losses which are computed between the source and converted utterances.
The losses facilitate the model to capture representations which are pertinent with respect to how humans perceive speech quality.
The models trained with the proposed losses produce less robotic voices compared to the baseline, and improves the overall quality for all target speakers. 

\begin{figure}[t]
     \centering

     \includegraphics[width=0.47\textwidth]{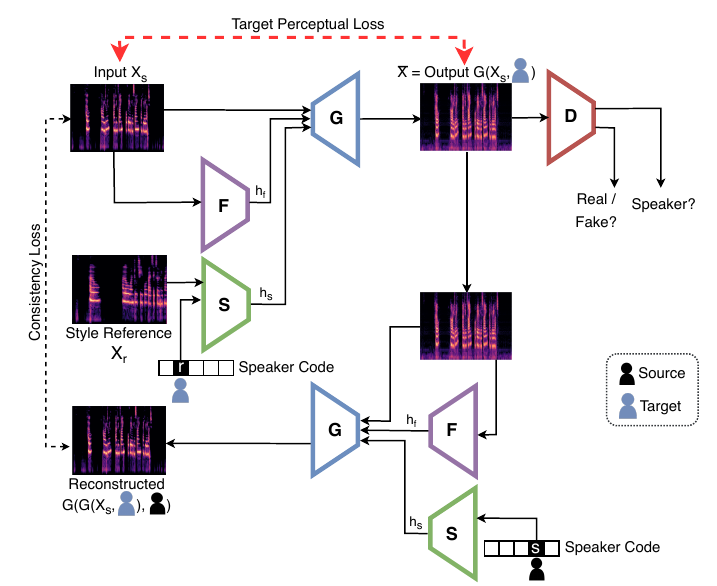}
\vspace{-1em}
\caption{StarGAN based Voice Conversion architecture}
\label{fig:starganv2c}
\end{figure}
\section*{Related Work}
In earlier VC approaches, parallel data was utilised, where utterances of the same linguistic content are available from both the source and target speakers.
%, which requires frame alignment. 
% The traditional parallel VC methods are based on parametric or non-parametric statistical modelling techniques
Many traditional statistical modelling-based parametric~\cite{Kobayashi2016TheNV,helander2011voice} and non-parametric~\cite{takashima2013exemplar,jin2016cute} parallel VC methods were proposed.
%, which use  to transform spectral features 
% The parametric ones such as Gaussian mixture model \cite{Kobayashi2016TheNV,stylianou1998continuous}, dynamic kernel partial least squares regression \cite{helander2011voice} and partial least square regression \cite{helander2010voice}, need more training data and are susceptible to over-smoothing than the non-parametric ones \cite{takashima2013exemplar,jin2016cute}.
%exemplar-based sparse representation techniques 
%The parametric ones are susceptible to over-smoothing and require more training data.
%compared to the non-parametric ones.
%time-domain discontinuity
Compared to the conventional methods, the sequence-to-sequence deep neural network (DNN) based method~\cite{liu2020voice} using parallel data produces less robotic voices.
However, it does not preserve prosody and produces mispronunciations~\cite{lian2020arvc}.
Further, the model learns one-to-one mapping, limiting its usage.

The recent works focus more on non-parallel data \cite{sisman2020overview}, as it is much easier and cheaper to collect.
% Some parallel approaches~\cite{silen2013voice,song2014text} were extended for non-parallel data using alignment techniques such as, iterative combination of a nearest neighbor search step and a conversion step alignment (INCA)~\cite{erro2009inca}.
A few VC methods~\cite{sun2016phonetic} use phonetic posteriorgrams (PPGs) as input to the encoder-decoder framework, which produces the translated acoustic features, consequently used by a vocoder to generate the converted speech.
The PPG-based conversions are generally not smooth resulting in degraded voice quality and naturalness ~\cite{sisman2020overview}.
Many non-parallel variational auto encoder (VAE)~\cite{sisman2020overview, wu2020one} methods were proposed, which typically disentangle the content and speaker embeddings using a reconstruction loss.
The VAE-based approaches are prone to spectrum smoothing, which leads to a buzzy-sounding voice ~\cite{sisman2020overview}.
%and time-domain discontinuity.
A plethora of GAN-based VC approaches \cite{kameoka2018stargan,sakamoto2021stargan,starganv2vc} have been proposed to overcome this over-smoothing effect.
%, where most methods perform any-to-many VC. %few \cite{eskimez2021one} any-to-any.   %where they were reported to produce naturally sounding voice compared to VAE approaches.
GAN-based VC approaches use cycle-consistency loss \cite{kaneko2018cyclegan}, which enable them to use non-parallel data.
\label{sec:rw}
\section*{Method}
\label{sec:method}
Our architecture is based on the GAN-based method StarGANv2-VC \cite{starganv2vc}. We describe the architecture and then the perceptual losses used in the work.

\subsection*{StarGANv2-VC}
StarGANv2-VC \cite{starganv2vc} is a non-parallel many-to-many voice conversion GAN-based approach.
The architecture is shown in Figure \ref{fig:starganv2c}.
% The method uses the cyclic consistency loss~\cite{kaneko2018cyclegan}, shown in Figure.
% In CycleGAN one generator is required for a unique pair of source and target speakers.
% Therefore, multiple generators are needed to generate the mapping between many-to-many speakers.
In StarGANv2-VC, only one generator is required for conversion among many pairs.
%, as shown in Figure \ref{fig:starganv2c}.
We describe the pertinent components of the framework, which are portrayed in Figure~\ref{fig:starganv2c}:
\begin{itemize}
\item \textbf{Generator}:
The generator G produces the converted mel-spectrogram $\overline{X}$ using three inputs: log mel-spectrogram generated from the source utterance $X_s$, fundamental frequency (F0) embeddings $h_f$ from the source utterance and target speaker's style-code $h_s$.
The F0 embeddings are the convolutional outputs from a pre-trained joint detection and classification (JDC) network \cite{kum2019joint}, which has convolutional layers followed by BLSTM units.
The converted mel-spectrogram $\overline{X}$ bears the style/timbre of the target speaker and the linguistic content of the source.

\item \textbf{Speaker Style Encoder}:
The speaker style encoder S captures representations of the speaker's style.
The style may represent accent, mannerism and other attributes which can be associated with the speaker independent of the content spoken.
Provided a mel-spectrogram $X_r$, which is generated from a reference utterance different from the source mel-spectrogram $X_s$ and a speaker-code $r$, the S generates the speaker style embeddings $h_s$.
The speaker-code is a one-hot encoding of the speaker labels.
The embedding $h_s$ acts as one of the inputs to the generator G, which contributes to the style of the converted mel-spectrogram $\overline{X}$.
S initially processes the input mel-spectrogram through multiple convolutional layers which are shared for all speakers, followed by a speaker-specific linear layer which maps the shared features into a speaker-specific style embedding.

\item \textbf{Discriminator and Speaker Classifier:} 
The architecture has a discriminator D, as present in any GAN model, which performs the quality check of the conversions by capturing the representations for the real and fake samples.
The additional adversarial speaker classifier (C) has the same architecture as D.
When the D is trained, keeping the weights of G fixed, the C classifies the source speaker, which encourages G to produce conversions having no trace of source speaker's attributes.
When G is trained, the weights of D are fixed, the C classifies the target speaker, which facilitates providing feedback to G, such that it produces conversions sounding like the target speaker.
\end{itemize}
\subsection*{Perceptual Losses}
%Much of the work in voice conversion is focused on architectural changes, using losses not penalising how humans perceive the degradation in voice quality.
Task specific perceptual losses facilitate models to capture pertinent representations, required to achieve the goal~\cite{ghosh2021perception}.
In our case, to improve the overall quality of voice conversions for all target speakers.
\label{sec:ploss}
\begin{itemize}
\item \textbf{Short Time Objective Intelligibility (STOI):} STOI \cite{stoi} is an intrusive metric that compares the degraded signal with the high quality ground truth to measure the intelligibility of the noisy signal.
The STOI score ranges from 0 to 1, with higher values indicating better intelligibility.
In our case, $X$ and $\bar{X}$ act as the ground truth and noisy signals respectively.
To calculate STOI, firstly speech signals are divided into short frames where each frame overlaps with the adjacent frames to capture the temporal context.
For each frame, the short-time power spectrum is calculated using a Fourier transform.
The modulation spectrum of both the signals are calculated by applying a perceptual auditory filter-bank (one-third octave band) to the short-time power spectra.
% The frequencies filtered by one-third octave band are the ones at which human ears are more sensitive in perceiving the difference in quality of signal.
% The modulation spectrum represents the distribution of energy in the speech signal across different modulation frequencies, which are critical with respect to human auditory perception.
The correlation coefficient between the modulation spectra of the original and degraded speech signals are calculated, which provides a similarity measure between two spectra.
The STOI score at time frame $m$ is calculated by taking average over all one-third octave bands as shown in Equation \ref{eqn: stoif}, where $j$ is the index of the one-third octave band, $x_{j,m}$ and $\bar{x}_{j,m}$ denote the vectors representing the short-term temporal envelopes for time frame m and one-third octave band $j$ of the clean and noisy signals respectively. $\mu$(·) denotes sample mean and $J$ is the total number of the one-third octave bands.
    \begin{equation}
    \label{eqn: stoif}
    f_{stoi}(X_m, \bar{X}_m) = \sum_{j=1}^J \frac{(x_{j,m} - \mu_{x_{j,m}}) (\bar{x}_{j,m} - \mu_{\bar{x}_{j,m}})}{\lVert x_{j,m} - \mu_{x_{j,m}} \rVert_{2} \lVert \bar{x}_{j,m} - \mu_{\bar{x}_{j,m}}\rVert_{2}}
\end{equation}
The loss is calculated as shown in Equation \ref{eq:stoiloss}, as done in \cite{stoi_loss}, where mean squared error (Equation \ref{eq:stoilossm}) is also considered along with STOI score (Equation \ref{eq:stoilosss}), as STOI calculates the discrepancy only for frequencies below 4.3 KHz.
$\lambda_{stoi}$ and $\lambda_{mse}$ are hyperparameters which weigh the contribution of $L_{stoi_s}$ and $L_{stoi_m}$ respectively.
\begin{equation}
\label{eq:stoilosss}
    L_{stoi_s} =  (1- f_{stoi}(X_m, \bar{X}_m))
\end{equation}
\begin{equation}
\label{eq:stoilossm}
L_{stoi_m} =  (\lVert X_m^{J} - \bar{X}_m^{J} \rVert_{1} / J)
\end{equation}
\begin{equation}
\label{eq:stoiloss}
    L_{stoi} = \frac{1}{m}(\lambda_{stoi} * L_{stoi_s} + \lambda_{mse} * L_{stoi_m})
\end{equation}
% \begin{equation}
%     d_{m} = f(X_m^{J}, Y_m^{J}) = \frac{1}{J}\sum_j d_{j,m}
% \end{equation}

    \item \textbf{Predicted Mean Opinion Score (pMOS):}
    MOS is a subjective measure which is used to assess the naturalness of the converted voice in voice conversion \cite{starganv2vc}.
    The measure correlates well with human perception of audio quality and naturalness.
    However, it is arduous and expensive as many participants' involvement is needed.
    Therefore, a measure similar to MOS is desirable, which captures the intrinsic naturalness of the conversions.
    MOSNet proposed in \cite{mosnet} can be used as a proxy MOS score generator.
    MOSNet is a combination of a convolutional neural network (CNN) and bidirectional long short-term memory (BLSTM) architecture.
    The CNN layers extract the representations required to assess the quality of the frames.
    BLSTM can effectively incorporate prolonged temporal dependencies and sequential traits into representative features.
    At the end two fully-connected layers are used, which regresses the frame-wise features into a frame-level quality score, which is followed by a global averaging operation to obtain the utterance-level score.
    The loss is calculated as shown in Equation \ref{eq:mossloss}, where MOS(.) denotes MOSNet and $\lambda_{mos}$ is a hyperparameter.
    The loss encourages G to produce conversions having naturalness similar to the original utterance.
    \begin{equation}
\label{eq:mossloss}
    L_{mos} = \lambda_{mos} * \lVert MOS(X) - MOS(\bar{X}) \rVert_1
\end{equation}
    \item \textbf{Pitch correlation coefficient (PCC):} Pitch is the perceptual measure of F0.
    The pitch contour contributes to the intonation or prosody of an utterance \cite{PCC}.
    PCC is the Pearson correlation between two normalised F0 contours, which provides the similarity between two utterances with respect to prosody \cite{PCC}.
    The F0 contours for two utterances having same content and intonation will vary for two groups (age, gender, etc).
    However, there should not be a large difference between the normalised F0 contours, i.e. the change in F0 over time should not vary much.
    Therefore, a higher PCC is desirable.
    PCC Loss is represented in Equation \ref{eq:pcc}, where Pearson(.) is the Pearson correlation operator.
    \begin{equation}
\label{eq:pcc}
    L_{pcc} = 1 - Pearson\Biggl(\frac{F(X)}{\lVert F(X) \rVert_1}, \frac{\bar{F}(X)}{\lVert \bar{F}(X) \rVert_1}\Biggr)
\end{equation}
\end{itemize}

\subsection*{Objective Function}
The generator G is trained with loss $L_G$, where $L_{adv}$ is the typical GAN adversarial loss, $L_{aspk}$ is the adversarial speaker classification loss, $L_{sty}$ is the style reconstruction loss and $L_{cyc}$ is the cyclic consistency loss, as proposed in \cite{starganv2vc}.
$L_{p}$ denotes one of the proposed perceptual losses.
\begin{equation}
\label{eqn:g}
L_G = \underset{G,S}{min} L_{adv} + \lambda_{aspk}L_{aspk} + \lambda_{sty}L_{sty} + \lambda_{cyc}L_{cyc} + \lambda_{p}L_{p}
\end{equation}
The discriminator D and classifier C are trained using the
objective function shown in Equation \ref{eqn:D}, where $L_{spk}$ is the speaker classification loss \cite{starganv2vc}.
%, and $\lambda_{spk}$ and $\lambda_{p}$ are the
%hyperparameters for the source speaker classification loss and perceptual loss respectively.
\begin{equation} \label{eqn:D}
    L_D = \underset{D, C}{min} -L_{adv} + \lambda_{spk}L_{spk} + \lambda_{p}L_{p}
\end{equation}
The $\lambda$ for the corresponding loss denotes the hyperparameter which weighs the loss's contribution.
\section*{Experiment Details}
We train all the models using English utterances of the 20 speakers from VCTK \cite{vctk} dataset, as done in \cite{starganv2vc}.
The utterances are resampled to 24 kHz and randomly split as 80\%/10\%/10\% (train/val/test).
The models are trained for 150 epochs and with batch size of 64.
The log-melspectrograms are derived from 2 second long utterances.
AdamW optimizer is used with initial learning rate of 0.0001.
%The speaker classifier contributes in training after 10 epochs.
The hyperparameters are set as: $\lambda_{spk}=0.1, \lambda_{aspk}=0.5, \lambda_{sty}=1, \lambda_{cyc}=1, \lambda_{stoi}=1, \lambda_{mse}=0.1$.
STOI is computed using hyperparameters same as in \cite{stoi}.
The naturalness of the conversions is evaluated using pMOS.
The intelligibility of the conversions is measured using character error rate (CER), using the transcriptions from Whisper \cite{whisper} medium-English model.
We use automatic speaker verification (ASV) to measure speaker similarity as done in \cite{starganv2vc}.
We trained an AlexNet as done in \cite{starganv2vc} for speaker classification for the selected twenty speakers.
The classification accuracy (Speaker CLS) serves as the objective metric to assess how close the conversions sound to the target speaker.

\section*{Results and Discussion}
\begin{figure}[t]
     \centering

     \includegraphics[width=0.47\textwidth]{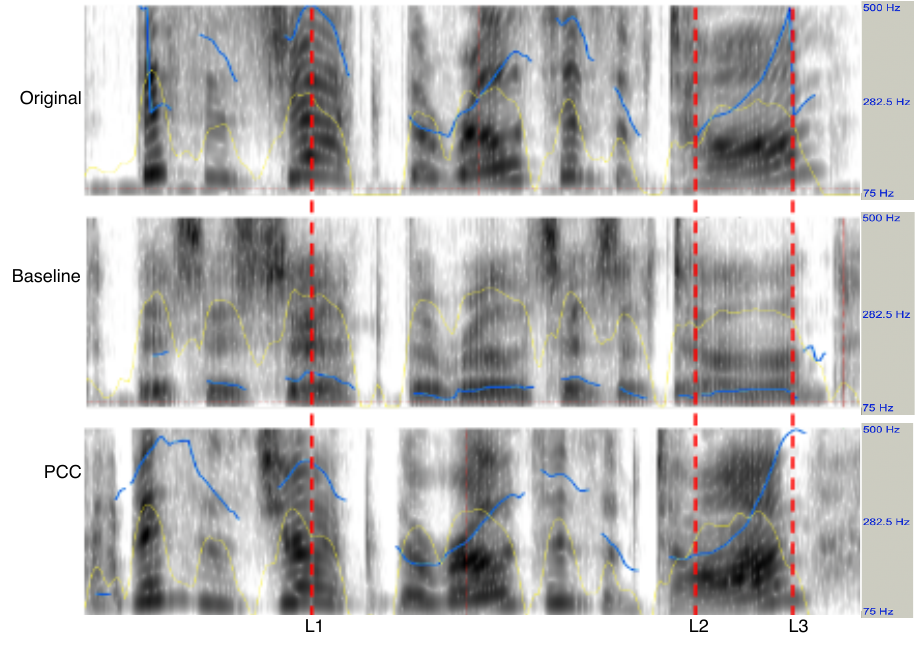}

\caption{Change in F0 contour (blue) for the original, conversions using baseline and PCC. The pictures are produced using Praat software, which also shows the F0 contour and intensity contour (yellow).}
\label{fig:praat}
\end{figure}

We randomly selected 5 male and 5 female speakers as the target speakers.
For each source speaker, randomly 50 utterances are selected, which leads to 1000 conversions.
The model trained using $L_{pcc}$ produces the best results with respect to naturalness and intelligibility, followed by the model trained using pMOS loss.
It is also observed the standard deviation for the baseline is much higher than the ones trained using target perceptual losses.
Therefore, the proposed losses produce better quality conversions overall, and not just for specific target speakers.
With respect to speaker similarity, all the models perform similarly, where PCC outperforms. 
\begin{table}[htbp]
    \centering
    \caption{Mean and standard deviation with objective metrics. Standard deviation in brackets.}
    \vspace{2mm}
    \label{tab:DescriptiveTextForATable}
        \begin{tabularx}{8cm}{@{\arrayrulewidth1.5pt\vline}Y@{\arrayrulewidth1.5pt\vline}Y|Y|Y@{\arrayrulewidth1.5pt\vline}}
            \noalign{\hrule height1.5pt} \multirow{2}{*}{Model} & \multicolumn{3}{c@{\arrayrulewidth1.5pt\vline}}{Objective Metrics}\\
            \cline{2-4} & MOS & CER & Speaker CLS\\
            \noalign{\hrule height1.5pt} Baseline & 3.20 (1.23) & 8.42 (9.39) & 78.7\%\\
            \hline STOI & 3.72 (0.32) & 3.01 (4.50) & 80.8\%\\
            \hline pMOS & 3.73 (0.23) & 3.10 (3.31) & 83.1\%\\
            \hline PCC & \textbf{3.80} (0.25) & \textbf{2.96} (3.78) & 85.8\%\\
            \noalign{\hrule height1.5pt}
        \end{tabularx}
\end{table}
It can also be seen that the models trained with perceptual losses maintain the intonation better than the baseline, as seen in Figure \ref{fig:praat}.
Further, it can been seen that in case of a large change in F0 contour (between L1 and L2), the baseline fails to maintain the structure.

\section*{Conclusion}
In this work we propose perceptual losses which are calculated between the original and converted samples conditioned on the target speaker.
The losses facilitate the model to disentangle the content and speaker representations, which leads to improved conversions not dependant on the target speaker selection.
In this work, we focus only on the naturalness and intelligibility aspect of voice quality.
As future work, we will perform listening tests to validate the results obtained through objective measures.
Further, we intend to incorporate perceptual losses which capture the emotional content as well.
This would be useful for the intelligent speech devices, whose response is driven by the emotion of the end-user.

\section*{Acknowledgements}
This research has been partly funded by the Federal Ministry of Education and Research of Germany in the project Emonymous (project number S21060A) and partly funded by the Volkswagen Foundation in the project AnonymPrevent (AI-based Improvement of Anonymity for Remote Assessment, Treatment and Prevention against Child Sexual Abuse).

\end{document}